\documentstyle[aps,amstex,epsfig,fleqn,floats]{revtex}
\begin{document}
%
%
%\draft
%\tighten
%
%
\title{Unconditionally Stable Algorithms to Solve the\\ Time-Dependent Maxwell Equations}
\author{Sebastiaan Kole, Marc Thilo Figge, and Hans De Raedt\\
%
%
%\address{
Centre for Theoretical Physics and Materials Science Centre \\
University of Groningen, Nijenborgh 4 \\
NL-9747 AG Groningen, The Netherlands \\
E-mail: j.s.kole@@phys.rug.nl, m.t.figge@@phys.rug.nl, deraedt@@phys.rug.nl\\
http://rugth30.phys.rug.nl/compphys
}
\date{\today}
\maketitle
%
%
%%%%%%%%%%%%%%%%%%%%%%%%%%%%%%%%%%%%%%%%%%%%%%%%%%%%%%%%%%%%%%%%%%%%%%%%%%%%%

\begin{abstract}
Based on the Suzuki product-formula approach, we construct a family of
unconditionally stable algorithms to solve the time-dependent Maxwell equations.
We describe a practical implementation of these algorithms for one-, two-, and three-dimensional
systems with spatially varying permittivity and permeability.
The salient features of the algorithms are illustrated
by computing selected eigenmodes and the full density of states
of one-, two-, and three-dimensional models and by simulating the propagation of
light in slabs of photonic band-gap materials.

\end{abstract}

%%%%%%%%%%%%%%%%%%%%%%%%%%%%%%%%%%%%%%%%%%%%%%%%%%%%%%%%%%%%%%%%%%%%%%%%%%%%%
%
%-------------------------------------------------------------
%
\section{Introduction}\label{sec1}
%
%-------------------------------------------------------------
%

%%%%%%%%%%%%%%%

Maxwell's partial differential equations of electromagnetism describe
the evolution of electric and magnetic fields in time and
space~\cite{BornWolf}.
They apply to a wide range of different physical situations that are
specified by the boundary conditions on the electromagnetic (EM) fields.
In many cases, numerical methods are required to solve Maxwell's
equations, either in the frequency or time domain.
For the time domain, a well-known class of algorithms is based on a
method proposed by Yee~\cite{Yee66} and is called the finite-difference
time-domain (FDTD) method.
The FDTD method has matured during past years and various algorithms
implemented for specific purposes are available by now~\cite{Taflove,Website}.
These algorithms owe their popularity mainly due to their flexibility and
speed while at the same time they are easy to implement.
A limitation of Yee-based FDTD techniques is that the stability of the algorithms
is conditional.
The stability depends on the mesh size used to discretize space and the time step
used to perform the time integration~\cite{Taflove}.

In this paper we present a family of unconditionally stable algorithms that solve
the time-dependent Maxwell equations (TDME) through the application of orthogonal transformations.
That this is possible follows from the representation of the TDME in matrix form.
The exponential of a skew-symmetric matrix plays the role of the time-evolution
operator of the EM fields.
This time-evolution operator is orthogonal.
The key to the construction of an unconditionally stable
algorithm to solve the Maxwell equations is the observation
that orthogonal approximations to this operator
automatically yield unconditionally stable algorithms.
The Lie-Trotter-Suzuki product formulae~\cite{Suzuki8591} provide
the mathematical framework to construct orthogonal approximations to
the time-evolution operator of the Maxwell equations.
However, this framework does not specify how to implement the algorithm.

Recently, a spectral-domain split-operator technique has been proposed to
solve the TDME~\cite{Harsh00}.
The split-operator approach is based on one of the many forms
of the Lie-Trotter-Suzuki product formulae.
The spectral-domain method makes use of Fast Fourier Transforms
to compute the matrix exponentials of the displacement operators.
The choice made in Ref.~\cite{Harsh00} yields an approximation to the time-evolution operator that
is no longer orthogonal and hence unconditional stability
is not automatically guarantueed~\cite{COMMENT1}.
In contrast, the methodology that we propose yields efficient, explicit,
unconditionally stable schemes that operate on the EM fields defined on the real space grid only.
This renders the algorithms rather flexible, avoids wrap-around effects~\cite{Harsh00},
and naturally allows for the spatial variations in both the permittivity and the permeability.
On the other hand, the implementation described in this paper is by no means unique
and leaves a lot of room for further improvements.

For EM fields in a homogeneous medium, Zheng et al. showed
that there is an alternating-direction-implicit (ADI) time-stepping algorithm
that is unconditionally stable \cite{Zheng1,Zheng2}.
Conceptually this approach is different from ours.
The Fourier-mode stability analysis performed
by Zheng et al.\cite{Zheng1} does not generalize to the case of
spatially varying permittivity and permeability whereas in our
approach, the algorithms are unconditionally stable by construction.

Our presentation is organized as follows:
The basic theoretical concepts are given in Sec.~\ref{sec2}.
In Sec.~\ref{sec3} we explain the general philosophy that underlies the
Suzuki approach to construct algorithms that are unconditionally stable.
In Sec.~\ref{sec4}, we show in detail how to implement
these ideas for the case of the TDME in one, two and three spatial dimensions, using algorithms
that are accurate up to second and fourth order in the time step.
In Sec.~\ref{sec5} we explain how we analyze the data generated by these algorithms.
In Sec.~\ref{sec6} we present the results of numerical
simulations for the physical systems that we selected as examples
to test the algorithms.
Our conclusions are given in Sec.~\ref{sec7}.

%
%-------------------------------------------------------------
%
\section{Theory}\label{sec2}
%
%-------------------------------------------------------------
%
\newcommand{\threevec}[3]{\left(\begin{array}{c}%
 #1 \\ #2 \\ #3 \end{array}\right)}
\newcommand{\ve}{\varepsilon}
\newcommand{\sve}{\sqrt{\varepsilon}}
\newcommand{\smu}{\sqrt{\mu}}
\def\bE{{\mathbf{E}}}
\def\bH{{\mathbf{H}}}
\def\bX{{\mathbf{X}}}
\def\bY{{\mathbf{Y}}}
\def\be{{\mathbf{e}}}
\def\br{{\mathbf{r}}}
\newcommand{\php}{\phantom{+}}
\newcommand{\pht}{\phantom{T}}
\newcommand{\dd}[1]{\frac{\partial}{\partial #1}}

The model system we consider in this paper describes EM fields in a $d$-dimensional ($d=1,2,3$) medium with spatially
varying permittivity and/or permeability,
surrounded by a perfectly conducting box.
In the absence of free charges and currents, the EM fields in such a
system satisfy Maxwell's equations~\cite{BornWolf}
\begin{eqnarray}
\dd{t}\,\bH & = & -\frac{1}{\mu} \nabla \times \bE\,,
\label{max1}\\[0.1cm]
\dd{t}\,{\bE} & = & \;\;\;\frac{1}{\ve} \nabla \times {\bH}\,,
\label{max2}\\[0.3cm]
{\rm div}\,\ve\bE & = & 0\,,
\label{max3}\\[0.3cm]
{\rm div}\,\bH & = & 0\,,
\label{max4}
\end{eqnarray}
where $\bH=(H_x(\br,t),H_y(\br,t),H_z(\br,t))^T$ and
$\bE=(E_x(\br,t),E_y(\br,t),E_z(\br,t))^T$ denote the magnetic and
electric field vector respectively.
The permeability and the permittivity are given
by $\mu=\mu(\br)$ and $\ve=\ve(\br)$.
For simplicity of notation, we will omit the spatial dependence on
$\br=(x,y,z)^T$ unless this leads to ambiguities.
On the surface of the perfectly conducting box the EM fields satisfy
the boundary conditions~\cite{BornWolf}
\begin{eqnarray}
\mathbf{n} \times \bE = 0 \quad ,\quad \mathbf{n} \cdot
\bH = 0, \label{boundcond}
\end{eqnarray}
with $\mathbf{n}$ denoting the vector normal to a boundary of the
surface.
The conditions Eq.~(\ref{boundcond}) assure that the normal component
of the magnetic field and the tangential components of the electric
field vanish at the boundary~\cite{BornWolf}.
Some important symmetries of the Maxwell equations
(\ref{max1})-(\ref{max4}) can be made explicit by introducing the
fields
\begin{eqnarray}
\bX(t)=\sqrt{\mu}\bH(t)\quad , \quad
\bY(t)=\sqrt{\varepsilon}\bE(t). \label{EY}
\end{eqnarray}
In terms of the fields $\bX(t)$ and $\bY(t)$, the TDME (Eqs.(\ref{max1}) and (\ref{max2})) read
\begin{equation} \dd{t} \left(\begin{array}{c} \bX(t) \\
\bY(t) \end{array} \right) =
 \left( \begin{array}{cc} 0 &
  -\frac{1}{\sqrt{\mu}}\mathbf{\nabla}\times\frac{1}
  {\sqrt{\varepsilon}} \\
   \frac{1}{\sqrt{\varepsilon}}\mathbf{\nabla}
   \times\frac{1}{\sqrt{\mu}} & 0
 \end{array}\right)
 \left(\begin{array}{c} \bX(t) \\ \bY(t)  \end{array} \right)
 \equiv {\mathcal H}
 \left(\begin{array}{c} \bX(t) \\ \bY(t)  \end{array} \right).
\label{eqn:mtxeqn}
\end{equation}
Writing $\Psi(t)=(\bX(t),\bY(t))^T$, Eq.~(\ref{eqn:mtxeqn}) becomes
\begin{equation}
\dd{t}\Psi(t)={\mathcal H}\Psi(t). \label{TDME}
\end{equation}
It is easy to show that ${\mathcal H}$ is skew-symmetric, i.e.
${\mathcal H}^T=-{\mathcal H}$, with respect to the inner product
\begin{equation}
\langle\Psi|\Psi^\prime\rangle\equiv\int_V\Psi^T\cdot\Psi^\prime\, d\br,
\label{INNER}
\end{equation}
where $V$ denotes the volume of the enclosing box.

The formal solution of Eq.~(\ref{TDME}) is given by
\begin{equation}
\Psi(t) = e^{t{\mathcal H}}\Psi(0), \label{eqn:formal}
\end{equation}
where $\Psi(0)$ represents the initial state of the EM fields and the
operator
\begin{equation}
U(t)= e^{t{\mathcal H}}, \label{Ut}
\end{equation}
determines their time evolution.
By construction
\begin{equation}
\|\Psi(t)\|^2 = \langle \Psi(t)|\Psi(t)\rangle =\int_V
\left[\varepsilon{\bE}^2(t) +\mu{\bH}^2(t)
\right] \, d{\br},
\label{normpsi}
\end{equation}
relating the length of $\Psi(t)$ to the energy density
\begin{equation}
w(t)\equiv
\varepsilon{\bE}^2(t) +\mu{\bH}^2(t),
\label{wt}
\end{equation}
of the EM fields~\cite{BornWolf}.
From $U(t)^T=U(-t)=U^{-1}(t)=e^{-t{\mathcal H}}$ it follows that
$\langle U(t)\Psi(0)|U(t)\Psi(0)\rangle=\langle\Psi(t)|\Psi(t)\rangle=
\langle\Psi(0)|\Psi(0)\rangle$.
Hence the time-evolution operator $U(t)$ is an orthogonal transformation,
rotating the vector $\Psi(t)$ without changing its length $\|\Psi\|$.
In physical terms this means that the energy density of the EM fields
does not change with time, as expected on physical grounds~\cite{BornWolf}.

The fact that $U(t)$ is an orthogonal transformation is essential for
the development of an unconditionally stable algorithm to solve the
Maxwell equations.
In practice, a numerical procedure solves the TDME by making use of an
approximation $\tilde U(t)$ to the true time evolution $U(t)$ (see below).
A necessary and sufficient condition for an algorithm to be
unconditionally stable is that~\cite{Smith85}
\begin{equation}
\|\tilde{U}(t)\Psi(0)\|\leq \|\Psi(0)\|.
\label{UNCOND}
\end{equation}
In other words, the length of $\Psi(t)$ should be bounded, for arbitrary
initial condition $\Psi(t=0)$ and for any time $t$~\cite{Smith85}.
By chosing for $\Psi(0)$ the eigenvector of $\tilde U(t)$ that corresponds
to the largest eigenvalue of $\tilde U(t)$, it follows from
Eq.~(\ref{UNCOND}) that the algorithm will be unconditionally stable by
construction if and only if the largest eigenvalue of $\tilde U(t)$
(denoted by $\|\tilde U(t)\|$) is less or equal than one~\cite{Smith85}.
If the approximation $\tilde U(t)$ is itself an orthogonal transformation, then
$\|\tilde U(t)\|=1$ and the numerical scheme will be unconditionally
stable.

Summarizing: Unconditionally stable algorithms to solve
Eq.~(\ref{eqn:mtxeqn}) can be constructed by employing orthogonal
approximations to the time-evolution $U(t)=e^{t{\mathcal H}}$.
For the case at hand, unconditional stability is tantamount to the exact
conservation of the energy density.

%
%-------------------------------------------------------------
%
\section{Unconditionally Stable Algorithms}\label{sec3}
%
%-------------------------------------------------------------
%

A numerical procedure that solves the TDME necessarily starts
by discretizing the spatial derivatives (see Sec.\ref{sec4}).
This maps the continuum problem described by ${\mathcal H}$ onto a
lattice problem defined by a matrix $H$.
Ideally, this mapping should not change the basic symmetries of the
original problem.
The underlying symmetry of the TDME suggests to use matrices $H$ that
are real and skew-symmetric.
Formally the time evolution of the EM fields on the lattice is given by
\begin{equation}
\Psi(t+\tau) = U(\tau)\Psi(t) = e^{\tau{H}}\Psi(t).
\label{TIMESTEP}
\end{equation}

The second ingredient of the numerical procedure is to choose an
approximation of the time-step operator $U(\tau)$.
A standard procedure is to truncate the Taylor series of the matrix
exponential~\cite{BELLMAN,GOLUB}
\begin{equation}
U(\tau)=e^{\tau H}=\sum_{n=0}^{\infty}\frac{(\tau{H})^n}{n!}.
\label{taylorU}
\end{equation}
Retaining terms up to first order in $\tau$ yields
\begin{equation}
\tilde{U}(\tau)=I+\tau{H},
\label{approxU0}
\end{equation}
where $I$ denotes the identity operator.
As $\tilde{U}(\tau)\tilde{U}(\tau)^T=I-(\tau{H})^2\not=I$ for
$\tau\not=0$, it is clear that the matrix Eq.~(\ref{approxU0}) is not
orthogonal.
Making use of the symmetry of $H$ and the positivity of the inner
product, we find that
$\langle\Psi(\tau)|\Psi(\tau)\rangle=
\langle\tilde{U}(\tau)\Psi(0)|\tilde{U}(\tau)\Psi(0)\rangle=
\langle\Psi(0)|\Psi(0)\rangle+\tau^2\langle H\Psi(0)|H \Psi(0)\rangle >
\langle\Psi(0)|\Psi(0)\rangle$,
implying that Eq.~(\ref{UNCOND}) does not hold.
Hence, according to the arguments given above, the (Euler) scheme Eq.~(\ref{approxU0}) is unstable,
a fact which is of course well-known~\cite{Smith85}.

The Yee algorithm is based on a leapfrog arrangement~\cite{Yee66} and
formally corresponds to an approximation to the matrix exponential $U(\tau)$
that can be written as
\begin{equation}
\tilde{U}_{\hbox{Yee}}(\tau)=I+\tau H_1+\tau^2 H_2\,,
\end{equation}
where $H_1$ and $H_2$ are matrices, the structure of which depend on the lattice dimensionality.
The presence of the second order contribution can render the algorithm stable
under certain conditions.
It seems difficult to determine these conditions for arbitrary (skew-symmetric)
$H_1$ and (symmetric) $H_2$, i.e. without making use of very specific
knowledge about the elements of $H_1$ and $H_2$.
For EM fields moving in free space, a Fourier-space stability analysis
of the Yee algorithm yields

\begin{equation}
\tau\leq\frac{\Delta}{c\sqrt{d}}\,,
\end{equation}
as the condition for stability~\cite{Taflove}.
Here $c$ is the light velocity in vacuum and $\Delta$ denotes the spatial mesh size~\cite{Taflove}.

A systematic approach to construct orthogonal approximations to matrix
exponentials, i.e. to construct unconditionally stable algorithms, is to
make use of the Lie-Trotter-Suzuki formula~\cite{Trotter59,Suzuki77}
\begin{equation}
e^{t(H_1+\ldots+H_p)}=
\lim_{m\rightarrow\infty}
\left(\prod_{i=1}^p e^{t{H}_i/m}\right)^m,
\label{TROT}
\end{equation}
and generalizations thereof~\cite{Suzuki8591,DeRaedt83}.
Applied to the case of interest here, the success of this approach relies
on the basic but rather trivial premise that the matrix $H$ can be written
as
\begin{equation}
{H}=\sum_{i=1}^{p}{H}_i,
\label{decompos}
\end{equation}
where each of the matrices ${H}_i$ is real and skew-symmetric.

The expression Eq.~(\ref{TROT}) suggests that
\begin{equation}
U_1(\tau)=e^{\tau{H}_1}\ldots e^{\tau{H}_p},
\label{tsapprox}
\end{equation}
might be a good approximation to $U(\tau)$ if $\tau$ is sufficiently small.
Most importantly, if all the $H_i$ are real and skew-symmetric, $U_1(\tau)$
is orthogonal by construction.
Therefore, by construction, a numerical scheme based on
Eq.~(\ref{tsapprox}) will be unconditionally stable.
Using the fact that both $U(\tau)$ and $U_1(\tau)$ are orthogonal matrices,
it can be shown that~\cite{DeRaedt87}
\begin{equation}
\|U(\tau)-U_1(\tau)\|\leq\frac{\tau^2}{2}\sum_{i<j}\|[{H}_i,{H}_j]\|\,,
\label{tserror}
\end{equation}
where $[{H}_i,{H}_j]=H_i H_j - H_j H_i$.
From Eq.~(\ref{tserror}) it follows that, in general, the Taylor series of
$U(\tau)$ and $U_1(\tau)$ are identical up to first order in $\tau$.
We will call $U_1(\tau)$ a first-order approximation to $U(\tau)$.

The product-formula approach provides simple, systematic procedures to
improve the accuracy of the approximation to $U(\tau)$ without changing its
fundamental symmetries.
For example the orthogonal matrix
\begin{equation}
U_2(\tau)={U_1(-\tau/2)}^TU_1(\tau/2)=
e^{\tau{H}_p/2}\ldots e^{\tau{H}_1/2}e^{\tau{H}_1/2}\ldots e^{\tau{H}_p/2},
\label{secordapp}
\end{equation}
is a second-order approximation to $U(\tau)$~\cite{Suzuki8591,DeRaedt83}.
Suzuki's fractal decomposition approach~\cite{Suzuki8591} gives a general
method to construct higher-order approximations based on $U_1(\tau)$ or
$U_2(\tau)$.
A particularly useful fourth-order approximation is given by~\cite{Suzuki8591}
\begin{equation}
U_4(\tau)=U_2(a\tau)U_2(a\tau)U_2((1-4a)\tau)U_2(a\tau)U_2(a\tau),
\label{fouordapp}
\end{equation}
where $a=1/(4-4^{1/3})$.
The approximations Eqs.(\ref{tsapprox}) and (\ref{secordapp}), and
(\ref{fouordapp}) have proven to be very useful in many applications
~\cite{Suzuki77,Chorin78,DeRaedt83,DeRaedt87,Koboyashi94,DeRaedt94,Rouhi95,Shadwick97,Krech98,Tran98,Michielsen98,DeRaedt00}
and, as we show below, turn out to be equally useful for solving the TDME.
In practice an efficient implementation of the first-order scheme is all
that is needed to construct the higher-order algorithms
Eqs.(\ref{secordapp}) and (\ref{fouordapp}).

To summarize: Suzuki's product-formula approach provides the formal
machinery to define algorithms that are unconditionally stable by
construction.
The accuracy of these algorithms can be improved systematically, to any
desired order~\cite{Suzuki8591}.
The only assumption made so far is that the real, skew-symmetric matrix $H$
representing the TDME can be written as a sum of $p$ real, skew-symmetric
matrices $H_i$.
The next step is to choose the $H_i$'s such that the matrix exponentials
$\exp(\tau H_1)$, ..., $\exp(\tau H_p)$ can be calculated efficiently.
This will turn the formal expressions for $U_2(\tau)$ and $U_4(\tau)$ into efficient
algorithms to solve the TDME.

%
%-------------------------------------------------------------
%
\section{Implementation}\label{sec4}
%
%-------------------------------------------------------------
%

In this section we present the details of our implementation of
unconditionally stable algorithms to solve the TDME based on the
Suzuki product-formula approach.
For pedagogical reasons we start by considering the simplest case:
A one-dimensional (1D) system.
Then we show that the strategy adopted for 1D readily extends to higher
spatial dimensions.
The implementation we describe below is by no means unique, leaving a lot
of room for further improvements.
In principle any decomposition Eq.~(\ref{decompos}) of $H$ into real
skew-symmetric parts will do.
Largely guided by previous work~\cite{DeRaedt87,DeRaedt94,Michielsen98,DeRaedt00},
we have adopted a decomposition that is
efficient, flexible, sufficiently accurate and easy to program.

%%%%%%%%%%%%%%%%%%%%%
\subsection{One dimension}\label{subsec41}
%%%%%%%%%%%%%%%%%%%%%

We consider a 1D system along the $x$-direction.
Accordingly, Maxwell's equations contain no partial derivatives with
respect to $y$ or $z$ and $\ve$ and $\mu$ do not depend on $y$ or $z$.
Under these conditions, the TDME reduce to two independent sets
of first-order differential equations~\cite{BornWolf}.
The solutions to these sets are known as the transverse electric (TE) mode
and the transverse magnetic (TM) mode~\cite{BornWolf}.
As the equations of the TE- and TM-mode only differ by a sign we can restrict
our considerations to the TM-mode and obtain the result for the TE-mode by
reversing the time.

From Eq.~(\ref{eqn:mtxeqn}) it follows that the magnetic field
$H_y(x,t)=X_y(x,t)/\sqrt{\mu(x)}$ and the electric field
$E_z(x,t)=Y_z(x,t)/\sqrt{\ve(x)}$ of the TM-mode are solutions of
\begin{eqnarray}
\dd{t}X_y(x,t) & = &
\frac{1}{\sqrt{\mu(x)}}\dd{x}\left(\frac{Y_z(x,t)}{\sqrt{\ve(x)}}\right),
\label{eqn:TM1Dx} \\
\dd{t}Y_z(x,t) & = &
\frac{1}{\sqrt{\ve(x)}}\dd{x}\left(\frac{X_y(x,t)}{\sqrt{\mu(x)}}\right).
\label{eqn:TM1Dy}
\end{eqnarray}
Note that in 1D, the divergence of $H_y(x,t)$ and $E_z(x,t)$ is zero.
Hence Eqs.(\ref{max3}) and (\ref{max4}) are automatically satisfied.

Using the second-order central-difference approximation to the first
derivative with respect to $x$, we obtain
\begin{eqnarray}
\dd{t}X_y(i,t) & = &
\frac{1}{\delta\sqrt{\mu_i}} \left(
\frac{Y_z(i+1,t)}{\sqrt{\varepsilon_{i+1}}}-
\frac{Y_z(i-1,t)}{\sqrt{\varepsilon_{i-1}}}\right),
\label{eqn:discrTMx} \\
\dd{t}Y_z(j,t) & = &
\frac{1}{\delta\sqrt{\varepsilon_j}} \left(
\frac{X_y(j+1,t)}{\sqrt{\mu_{j+1}}}-
\frac{X_y(j-1,t)}{\sqrt{\mu_{j-1}}}\right),
\label{eqn:discrTMy}
\end{eqnarray}
where the integer $i$ labels the grid points and $\delta$ denotes the
distance between two next-nearest neighbor lattice points (hence the
absence of a factor two in the nominator).
For notational simplicity we will, from now on, specify the spatial
coordinates through the lattice index $i$, e.g. $X_y(i,t)$ stands for $X_y(x=(i+1)\delta/2,t)$.

\begin{figure}[ht]
\setlength{\unitlength}{1cm}
\begin{picture}(8.0,1.6)
\put(0.5,0){\epsfig{file=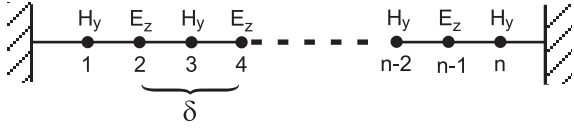}}
\end{picture}
\caption{Positions of the two TM-mode field components on the one-dimensional grid.}
\label{fig:fig1}
\end{figure}
Following Yee~\cite{Yee66} it is convenient to assign $X_y(i,t)$ and
$Y_z(j,t)$ to the odd, respectively, even numbered lattice site, as shown
in Fig.~\ref{fig:fig1} for a grid of $n$ points.
The equations (\ref{eqn:discrTMx}) and (\ref{eqn:discrTMy}) can now be
combined into one equation of the form Eq.~(\ref{TDME}) by introducing
the $n$-dimensional vector
\begin{equation}
\Psi(i,t) =
\left\{ \begin{array}{lll} X_y(i,t)=\sqrt{\mu_i}H_y(i,t), & \mbox{$i$ odd}
\\ Y_z(i,t)=\sqrt{\varepsilon_i}E_y(i,t), & \mbox{$i$ even}
\end{array} \right..
\label{eqn:eqnindx}
\end{equation}
The vector $\Psi(t)$ describes both the magnetic and the electric field
on the lattice points $i=1,\ldots,n$.
As usual the $i$-th element of $\Psi(t)$ is given by the inner product
$\Psi(i,t)=\be^T_i\cdot\Psi(t)$ where $\be_i$ denotes the $i$-th unit vector
in the $n$-dimensional vector space.
Using this notation, it is easy to show that Eqs.(\ref{eqn:discrTMx}) and
(\ref{eqn:discrTMy}) reduce to
\begin{equation}
\dd{t}\Psi(t)=H \Psi(t),
\end{equation}
where the matrix $H$ is given by
\begin{equation}
H=
\mathop{{\sum}'}_{i=1}^{n} \left[
\beta_{i+1,i}
\left(
\be^{\pht}_{i}\be^T_{i+1}-
\be^{\pht}_{i+1}\be^T_{i}\right)+
\beta_{i+1,i+2}
\left(
\be^{\pht}_{i+1}\be^T_{i+2}-
\be^{\pht}_{i+2}\be^T_{i+1}\right)\right]\,,
\label{eqn:oper_1D}
\end{equation}
with $\beta_{i,j}=1/(\delta\sqrt{\varepsilon_i \mu_j})$ and the prime
indicates that the sum is over odd integers only.
In complete analogy to Eq.~(\ref{eqn:formal}) the time evolution of $\Psi(t)$
is formally given by $\Psi(t) = U(t) \Psi(0)$ with $U(t) =\exp(t{H})$.

The notation introduced above will prove most useful for the case of 2D and
3D for which it is rather cumbersome to write down matrix representations.
For the 1D case it is not difficult and in fact very instructive to write
down the matrix $H$ explicitly.
Indeed, we have
\begin{equation}
H=
%\frac{1}{\delta}
\left( \begin{array}{ccccc}
0 & \beta_{2,1} & & & \\
-\beta_{2,1} & 0 & \beta_{2,3} & & \\
& \ddots & \ddots & \ddots & \\
& & -\beta_{n-1,n-2} & 0 & \beta_{n-1,n} \\
& & & -\beta_{n-1,n} & 0
\end{array} \right),
\label{H}
\end{equation}
and we immediately see that $H$ is skew-symmetric by construction.
Furthermore, for $n$ odd we have
\begin{equation}
\dd{t}\Psi(1,t)=\beta_{2,1}
\Psi(2,t)\;\;\;\;\;{\rm and}\;\;\;\;\;
\dd{t}\Psi(n,t)=-\beta_{n-1,n}
\Psi(n-1,t)\,,
\end{equation}
such that the electric field vanishes at the boundaries
($Y_z(0,t)=Y_z(n+1,t)=0$, see also Fig.~\ref{fig:fig1}), as required by
the boundary conditions Eq.~(\ref{boundcond}).
For this reason we only consider the case of $n$ odd in the sequel.

According to the general procedure outlined in Sec.\ref{sec3}, the final step in the
construction of an unconditionally stable algorithm is to decompose $H$.
Guided by previous work on Schr\"odinger and diffusion
problems~\cite{DeRaedt87,DeRaedt94,Michielsen98,DeRaedt00}, we split
$H$ into two parts, i.e. $H=H_{1}+H_{2}$, where
\begin{eqnarray}
H_{1} &=&
\mathop{{\sum}'}_{i=1}^{n}
\beta_{i+1,i}\left(
\be^{\pht}_{i}\be^T_{i+1}-
\be^{\pht}_{i+1}\be^T_{i}
\right),
\label{h1}
\\
H_{2} &=&
\mathop{{\sum}'}_{i=1}^{n}
\beta_{i+1,i+2}\left(
\be^{\pht}_{i+1}\be^T_{i+2}-
\be^{\pht}_{i+2}\be^T_{i+1}
\right),
\label{h2}
\end{eqnarray}
i.e. we divide the lattice into odd and even numbered cells.
In matrix notation we have
\begin{equation}
H_{1} =
\left( \begin{array}{cccccccc}
0         & \beta_{2,1} &       &        &        &           &        & \\
-\beta_{2,1} & 0      & 0       &        &        &           &        & \\
        & 0      & 0       & \beta_{4,3} &        &           &        & \\
        &        & -\beta_{4,3} & 0      & 0      &           &        & \\
        &        &         & 0           & \ddots & 0         &        & \\
        &        &         &        & 0  & 0         & \beta_{n-1,n-2} & \\
        &        &         &        &    & -\beta_{n-1,n-2}   & 0      & 0 \\
        &        &         &        &    &                    & 0      & 0
\end{array} \right)
\label{h1m}
\end{equation}
and
\begin{equation}
H_{2} =
\left( \begin{array}{cccccccc}
0 & 0      &         &        &           &          &           & \\
0 & 0      & \beta_{2,3}  &   &           &          &           & \\
  & -\beta_{2,3}& 0       & 0 &           &          &           & \\
  &        & 0       & \ddots & 0         &          &           & \\
  &        &         & 0      & 0         & \beta_{n-3,n-2} &     & \\
  &        &         &        & -\beta_{n-3,n-2} & 0        & 0   & \\
  &        &         &        &           & 0        & 0  & \beta_{n-1,n} \\
  &        &         &        &           &          & -\beta_{n-1,n} & 0
\end{array} \right).
\label{h2m}
\end{equation}
Clearly both $H_{1}$ and $H_{2}$ are skew-symmetric block-diagonal matrices,
containing one $1\times1$ matrix and $(n-1)/2$ real, $2\times2$
skew-symmetric matrices.

According to the general theory given above, the first-order algorithm defined by
\begin{equation}
U_1(\tau)=e^{\tau H_{1}} e^{\tau H_{2}}.
\label{U1}
\end{equation}
is all that is needed to construct unconditionally stable second and higher-order algorithms.
As the matrix exponential of a block-diagonal matrix is equal to the
block-diagonal matrix of the matrix exponentials of the individual blocks,
the numerical calculation of $e^{\tau H_{1}}$ (or $e^{\tau H_{2}}$) reduces
to the calculation of $(n-1)/2$ matrix exponentials of $2\times2$ matrices.
The matrix exponential of a typical $2\times2$ matrix appearing in
$e^{\tau H_{1}}$ or $e^{\tau H_{2}}$ is given by
\begin{eqnarray}
% \exp\left[\alpha\left(
% \be^{\pht}_{i}\be^T_{j}-
% \be^{\pht}_{j}\be^T_{i}
% \right)\right]
% \left(\begin{array}{c} \Psi(i,t) \\ \Psi(j,t) \end{array} \right)
% &=&
\exp\left[\alpha
\left(\begin{array}{cc} \php0&1\\ -1&0\end{array}\right)\right]
\left(\begin{array}{c} \Psi(i,t) \\ \Psi(j,t) \end{array} \right)
\label{twobytwo} % \\
&=&\left(
\begin{array}{cc}
\php\cos \alpha & \php\sin\alpha
\\ -\sin \alpha & \php\cos\alpha
\end{array}\right)
\left(\begin{array}{c} \Psi(i,t) \\ \Psi(j,t) \end{array} \right).
\end{eqnarray}

%%%%%%%%%%%%%%%%%%%%%
\subsection{Two dimensions}\label{subsec42}
%%%%%%%%%%%%%%%%%%%%%

Assuming translational invariance with respect to the $z$ direction,
the system effectively becomes 2D and the TDME
separate into two sets of equations~\cite{BornWolf}.
For conciseness, in this section we only discuss the set of equations for
the TM-modes.
The TE-modes can be treated in exactly the same manner.

\begin{figure}[h]
\setlength{\unitlength}{1cm}
\begin{picture}(9.15,5.8)
\put(0.5,0){\epsfig{file=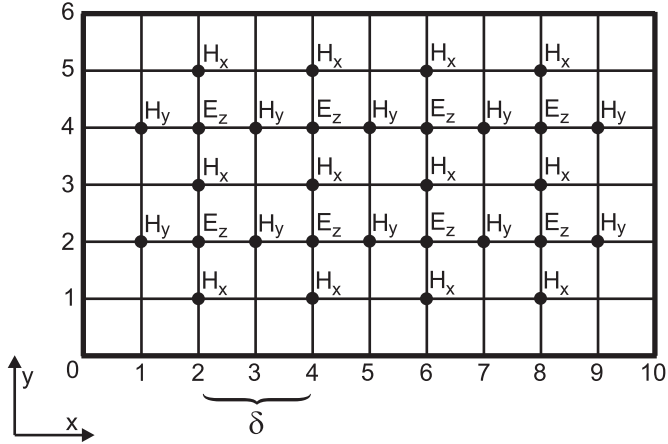}}
\end{picture}
\caption{Positions of the three TM-mode EM-field components on the two-dimensional grid
for $n_x=9$ and $n_y=5$.}
\label{fig:2DTM}
\end{figure}

The relevant EM fields for the
TM-modes in 2D are $\Psi(t)=(\bX_x(x,y,t),\bX_y(x,y,t),\bY_z(x,y,t))^T$, in terms of which
TDME Eq.(\ref{eqn:mtxeqn}) reads
\begin{equation}
\dd{t}\Psi(t) = {\mathcal H}\Psi(t)=
\left(\begin{array}{ccc} 0 & 0 &
          -\frac{1}{\smu}\dd{y}\frac{1}{\sve} \\ 0 & 0 &
\php\frac{1}{\smu}\dd{x}\frac{1}{\sve} \\
-\frac{1}{\sve}\dd{y}\frac{1}{\smu} &
\php\frac{1}{\sve}\dd{x}\frac{1}{\smu} & 0 \end{array}\right)
\Psi(t)\,. \label{eqn:2DTM}
\end{equation}
We discretize continuum space by simply re-using the one-dimensional lattice
introduced above, as exemplified in Fig.~\ref{fig:2DTM} for the case of the
TM-modes.
This construction automatically takes care of the boundary conditions if
$n_x$ and $n_y$ are odd and yields a real skew-symmetric matrix $H$.
In analogy with the 1D case the elements of $\Psi(t)$ are defined by
\begin{equation}
\Psi(i,j,t) = \left\{ \begin{array}{lll} Y_z(i,j,t)=\sqrt{\varepsilon_{i,j}}E_z(i,j,t), & \mbox{$i$
even and $j$ even} \\ X_y(i,j,t)=\sqrt{\mu_{i,j}}H_y(i,j,t), & \mbox{$i$ odd and $j$ even}
\\ X_x(i,j,t)=\sqrt{\mu_{i,j}}H_x(i,j,t), & \mbox{$i$ and $j$ odd}
\end{array} \right.\,.
\end{equation}
Discretization of the differential operators that appear in
Eq.~(\ref{eqn:2DTM}) yield expressions that have the same structure as
Eq.~(\ref{H}), with extra subscripts to account for the second spatial
dimension.
It follows that on the lattice,
\begin{equation}
\dd{t}\Psi(t)={H} \Psi(t)= \mathop{{\sum}'}_{i=1}^{n_x}
\mathop{{\sum}'}_{j=1}^{n_y} \left[ H^{(x)}(i,j)+H^{(y)}(i,j)
\right]\Psi(t), \label{2Dsplit}
\end{equation}
where
\begin{eqnarray}
H^{(x)}(i,j)=&+& \frac{\be^{\pht}_{i,j+1}\be^T_{i+1,j+1}-
\be^{\pht}_{i+1,j+1}\be^T_{i,j+1}}
{\delta\sqrt{\ve_{i+1,j+1}\mu_{i,j+1}}} +
\frac{\be^{\pht}_{i+1,j+1}\be^T_{i+2,j+1}-
\be^{\pht}_{i+2,j+1}\be^T_{i+1,j+1}}
{\delta\sqrt{\ve_{i+2,j+1}\mu_{i+2,j+1}}}\,,
\label{eqn:2Dsplit_discr_x}
\\
H^{(y)}(i,j)=&-& \frac{\be^{\pht}_{i+1,j}\be^T_{i+1,j+1}-
\be^{\pht}_{i+1,j+1}\be^T_{i+1,j}}
{\delta\sqrt{\ve_{i+1,j+1}\mu_{i+1,j}}} -
\frac{\be^{\pht}_{i+1,j+1}\be^T_{i+1,j+2}-
\be^{\pht}_{i+1,j+2}\be^T_{i+1,j+1}}
{\delta\sqrt{\ve_{i+1,j+2}\mu_{i+1,j+1}}}\,,
\label{eqn:2Dsplit_discr_y}
\end{eqnarray}
and the superscripts $(x)$ and $(y)$ refer to the derivative with respect to
$x$ and $y$ respectively.
Note that we use the pair $(i,j)$ to label the $n_xn_y$ unit vectors
$\be_{i,j}$.
In complete analogy with the 1D case we split
Eqs.(\ref{eqn:2Dsplit_discr_x}) and (\ref{eqn:2Dsplit_discr_y}) into
two parts and obtain for the first-order approximation to $U(\tau)$
\begin{eqnarray}
U_1(\tau) & = & e^{\tau{H_{1}^{(x)}}} e^{\tau{H_{2}^{(x)}}}
e^{\tau{H_{1}^{(y)}}} e^{\tau{H_{2}^{(y)}}}\,, \label{U1D2}
\end{eqnarray}
where for instance, in formal analogy to Eq.~(\ref{h2}), we have
\begin{equation}
H_2^{(x)}\,=\,\mathop{{\sum}'}_{i=1}^{n_x}
\mathop{{\sum}'}_{j=1}^{n_y}
\frac{\be^{\pht}_{i+1,j+1}\be^T_{i+2,j+1}-
\be^{\pht}_{i+2,j+1}\be^T_{i+1,j+1}}
{\delta\sqrt{\ve_{i+2,j+1}\mu_{i+2,j+1}}}\,.
\end{equation}
It is not difficult to convince oneself that
approximation $U_1(\tau)$ and hence also $U_2(\tau)$ and $U_4(\tau)$
do not commute with the (lattice version of) the divergence.
Therefore the divergence of the EM fields in 2D is not conserved.
Although the initial state (at $t=0$) of the EM fields satifies
Eqs.(\ref{max3}) and (\ref{max4}), time-integration of the TDME
using $U_k(\tau)$ yields a solution that will not satisfy Eqs.(\ref{max3}) and (\ref{max4}).
However, for algorithm $U_k(\tau)$ the deviations from zero vanish as $\tau^k$
so that in practice these errors are under control and can be made
sufficiently small for practical purposes.

%%%%%%%%%%%%%%%%%%%%%
\subsection{Three dimensions}\label{subsec43}
%%%%%%%%%%%%%%%%%%%%%

\begin{figure}[ht]
\setlength{\unitlength}{1cm}
\begin{picture}(7,6.2)
\put(0.5,0){\epsfig{file=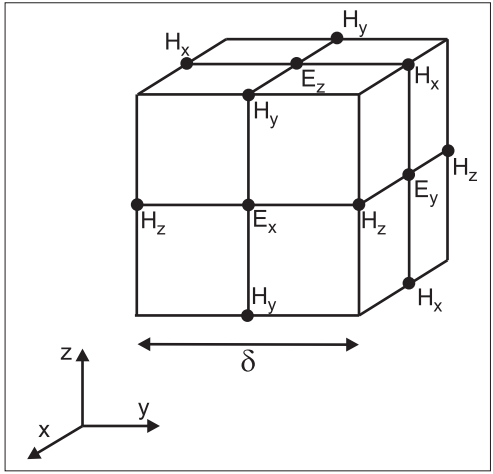}}
\end{picture}
\caption{Unit cell of the Yee grid.}
\label{fig:3D}
\end{figure}

In terms of $\Psi(t)=(\bX(t),\bY(t))^T$ for a 3D system,
Eq.~(\ref{eqn:mtxeqn}) reads
\begin{equation}
\dd{t} \Psi(t) = {\mathcal H} \Psi(t) =
\left(\begin{array}{cc} \php0^{\phantom{T}} &
\php{h}^{\phantom{T}} \\
-h^T & \php 0^{\phantom{T}} \end{array}\right)\Psi(t),\label{eqn:mtx3D}
\end{equation}
where $h$ is given by
\begin{equation}
h=\left(\begin{array}{ccc}
0 & \php\frac{1}{\smu}\dd{z}\frac{1}{\sve}
& -\frac{1}{\smu}\dd{y}\frac{1}{\sve} \\
-\frac{1}{\smu}\dd{z}\frac{1}{\sve} & 0 &
\php\frac{1}{\smu}\dd{x}\frac{1}{\sve} \\
\php\frac{1}{\smu}\dd{y}\frac{1}{\sve} &
-\frac{1}{\smu}\dd{x}\frac{1}{\sve} & 0
\end{array}\right).
\label{h}
\end{equation}
We discretize the spatial coordinates by adopting the standard Yee
grid~\cite{Yee66}.
We show a unit cell of this grid in Fig.~\ref{fig:3D}.

\noindent
In analogy to the systems in 1D and 2D, we assign the EM fields to the
lattice points such that the boundary conditions Eq.~(\ref{boundcond})
are automatically satisfied.
The elements of the vector $\Psi(i,j,k,t)$ are given by
\begin{equation}
\Psi(i,j,k,t) = \left\{ \begin{array}{lll} X_x(i,j,k,t)=\sqrt{\mu_{i,j,k}}H_x(i,j,k,t), &
\mbox{$i$ even, $j$ odd, $k$ odd}
\\ X_y(i,j,k,t)=\sqrt{\mu_{i,j,k}}H_y(i,j,k,t), & \mbox{$i$ odd, $j$
even, $k$ odd} \\ X_z(i,j,k,t)=\sqrt{\mu_{i,j,k}}H_z(i,j,k,t), & \mbox{$i$ odd, $j$ odd, $k$
even} \\ Y_x(i,j,k,t)=\sqrt{\varepsilon_{i,j,k}}E_x(i,j,k,t), & \mbox{$i$ odd, $j$ even, $k$ even}
\\ Y_y(i,j,k,t)=\sqrt{\varepsilon_{i,j,k}}E_y(i,j,k,t), & \mbox{$i$ even, $j$ odd, $k$ even} \\
Y_z(i,j,k,t)=\sqrt{\varepsilon_{i,j,k}}E_z(i,j,k,t), & \mbox{$i$ even, $j$ even, $k$ odd}
\\
\end{array} \right.\,,
\label{psidef}
\end{equation}
for the origin of the coordinate system $(i,j,k)=(0,0,0)$ at the center of
the unit cell shown in Fig.~\ref{fig:3D}.
The number of lattice points in the $x$, $y$, and $z$ direction will be
denoted by $n_x$, $n_y$, and $n_z$ respectively. As before these numbers
are assumed to be odd.

Discretization of the differential operators that appear in Eq.~(\ref{h})
yields Eq.~(\ref{eqn:mtxeqn}) in the form
\begin{equation}
\dd{t}\Psi(t)={H} \Psi(t)=
\mathop{{\sum}'}_{i=1}^{n_x}
\mathop{{\sum}'}_{j=1}^{n_y}
\mathop{{\sum}'}_{k=1}^{n_z}
\left[
H^{(x)}(i,j,k)+H^{(y)}(i,j,k)+H^{(z)}(i,j,k)
\right]\Psi(t),
\label{3Dsplit}
\end{equation}
where the superscripts $(x)$, $(y)$ and $(z)$ refer to the derivative
with respect to $x$, $y$ and $z$ respectively,
\begin{eqnarray}
H^{(x)}(i,j,k)=&+&
\frac{\be^{\pht}_{i,j+1,k}\be^T_{i+1,j+1,k}-
\be^{\pht}_{i+1,j+1,k}\be^T_{i,j+1,k}}
{\delta\sqrt{\ve_{i+1,j+1,k}\mu_{i,j+1,k}}}
-
\frac{\be^{\pht}_{i,j,k+1}\be^T_{i+1,j,k+1}-
\be^{\pht}_{i+1,j,k+1}\be^T_{i,j,k+1}}
{\delta\sqrt{\ve_{i+1,j,k+1}\mu_{i,j,k+1}}}
\label{hx} \\
&+&
\frac{\be^{\pht}_{i+1,j+1,k}\be^T_{i+2,j+1,k}-
\be^{\pht}_{i+2,j+1,k}\be^T_{i+1,j+1,k}}
{\delta\sqrt{\ve_{i+1,j+1,k}\mu_{i+2,j+1,k}}}
-
\frac{\be^{\pht}_{i+1,j,k+1}\be^T_{i+2,j,k+1}-
\be^{\pht}_{i+2,j,k+1}\be^T_{i+1,j,k+1}}
{\delta\sqrt{\ve_{i+1,j,k+1}\mu_{i+2,j,k+1}}},
\nonumber
\end{eqnarray}
and the expressions for $H^{(y)}(i,j,k)$ and $H^{(z)}(i,j,k)$ follow from
Eq.~(\ref{hx}) by symmetry.
Note that we use the triple $(i,j,k)$ to label the $n_xn_yn_z$ unit vectors
$\be_{i,j,k}$.

In complete analogy with the 1D and 2D case we split Eq.~(\ref{hx}) (as well
as $H^{(y)}(i,j,k)$ and $H^{(z)}(i,j,k)$) in two parts and obtain for the
first-order approximation to $U(\tau)$
\begin{eqnarray}
U_1(\tau) & = &
e^{\tau{H_{1}^{(x)}}}
e^{\tau{H_{2}^{(x)}}}
e^{\tau{H_{1}^{(y)}}}
e^{\tau{H_{2}^{(y)}}}
e^{\tau{H_{1}^{(z)}}}
e^{\tau{H_{2}^{(z)}}}\,,
\label{U1D3}
\end{eqnarray}
where for instance
\begin{equation}
H_{1}^{(z)} =
\mathop{{\sum}'}_{i=1}^{n_x}
\mathop{{\sum}'}_{j=1}^{n_y}
\mathop{{\sum}'}_{k=1}^{n_z}
\left(
\frac{\be^{\pht}_{i,j+1,k}\be^T_{i,j+1,k+1}-
\be^{\pht}_{i,j+1,k+1}\be^T_{i,j+1,k}}
{\delta\sqrt{\ve_{i,j+1,k+1}\mu_{i,j+1,k}}}-
\frac{\be^{\pht}_{i+1,j,k}\be^T_{i+1,j,k+1}-
\be^{\pht}_{i+1,j,k+1}\be^T_{i+1,j,k}}
{\delta\sqrt{\ve_{i+1,j,k+1}\mu_{i+1,j,k}}}
\right).
\label{h1z}
\end{equation}
Note that each contribution to Eq.~(\ref{h1z}) acts on a different pair of
elements of $\Psi(t)$.
Hence each of the matrix exponentials in Eq.~(\ref{U1D3}) acts on one quarter
of all the lattice points.
Performing the time-step operation Eq.~(\ref{U1D3}) involves only two sweeps
over all lattice points.

By construction the algorithm defined by Eq.~(\ref{U1D3}) is unconditionally
stable and so are the higher-order algorithms defined by  $U_2(\tau$) and
$U_4(\tau)$.
Each contribution to e.g. Eq.~(\ref{h1z}) is of the form Eq.~(\ref{h1}) and
hence its matrix exponential can be calculated in exactly the same manner as
in the 1D case (see Eq.~(\ref{twobytwo})).
The divergence of the EM fields in 3D is, for the same
reason as in the 2D case, not conserved but decreases as $\tau^k$.

\subsection{Implementation: Summary}\label{subsec44}

The notation required to write down the algorithms in mathematical
form might give the impression that these algorithms are difficult to program.
Actually that is not the case, on the contrary.
Recall that the first-order algorithm $U_1(\tau)$ is all we need to program:
As explained in Sec.~\ref{sec3}, more accurate schemes can be implemented
without extra programming.
Let us consider the algorithm for the case of 1D. For 2D and 3D we simply repeat the steps
described below two, respectively, three times.
We have
\begin{eqnarray}
U_1(\tau)&=&e^{\tau H_{1}} e^{\tau H_{2}}
\nonumber \\
&=&
\left\{\mathop{{\prod}'}_{i=1}^{n}
\exp\left[\beta_{i+1,i}\left(
\be^{\pht}_{i}\be^T_{i+1}-
\be^{\pht}_{i+1}\be^T_{i}
\right)\right]\right\}
\left\{\mathop{{\prod}'}_{i=1}^{n}
\exp\left[\beta_{i+1,i+2}\left(
\be^{\pht}_{i+1}\be^T_{i+2}-
\be^{\pht}_{i+2}\be^T_{i+1}
\right)\right]\right\}.
\label{U1D}
\end{eqnarray}
where we used the block-diagonal structure of $H_1$ and $H_2$ (see Eqs.(\ref{h1m}) and (\ref{h2m}))
to obtain an exact expression for $U_1(\tau)$ in terms of an ordered
product of matrix exponentials.
Each of these matrix exponentials only operates on a pair of elements of $\Psi(t)$
and leaves other elements intact.
The indices of each of these pairs are given by the subscripts of $\be$ and
$\be^T$.
From Eq.(\ref{U1D}) it is then clear what the program should do:
Make loops over $i$ with stride 2.
For each $i$ pick a pair of elements from $\Psi(t)$ according to the
subscripts of $\be$ and $\be^T$, compute (or fetch from memory) the elements of the plane rotation
(see Eq.~(\ref{twobytwo})), perform the plane rotation, i.e. multiply the
$2\times2$ matrices and the vectors of length two, and overwrite the same
two elements.

It also follows immediately that performing a time step with algorithms
based on Eq.~(\ref{U1D}) takes ${\cal O}(K)$ plane rotations where
$K$ is the total number of elements of the vector $\Psi(t)$
(which is less or equal to the number of grid points).
This renders the algorithm efficient:
The number of operations to complete one time step scales linearly with the
number of grid points.
Also note that there is a high degree of intrinsic parallelism in this class
of algorithms.
In principle, the $(n-1)/2$ matrix-vector multiplications that implement
$e^{\tau H_{1}}$ or $e^{\tau H_{2}}$ can be done in parallel.

In the absence of external currents (see below), updating
the EM field values of a 3D system using the the Yee algorithm requires 6 arithmetic operations
(see Eq.(33) in Ref.\cite{Taflove})
whereas the second-order algorithm $U_2(\tau)$ requires 33 arithmetic operations.
For 1D and 2D problems the ratio is 9/4 and 21/6 respectively.
Thus, in terms of CPU time, the price paid for the unconditional stability
of the algorithms is not that much and for some applications (see below)
may well be worth paying.

An important aspect that we have not yet discussed is the effect of the
discretization of space on the accuracy of the numerical results.
Both conditional Yee-type algorithms and
unconditionally stable algorithms $U_1(\tau)$, $U_2(\tau)$ and $U_4(\tau)$
suffer from numerical dispersion (see Ref.\cite{Taflove}, chapter 4 for an in-depth
discussion). Simple methods to reduce numerical dispersion are
taking a finer mesh or employing more accurate finite-difference approximations
for the spatial derivates \cite{Taflove}.
The former obviously can be used here too
(for the simulations discussed below we used a mesh size that yields sufficiently
accurate results for the present purposes).
There are no fundamental nor practical problems to incorporate
the latter method in the Suzuki-product-formula approach \cite{DeRaedt87,DeRaedt94}.
However as the emphasis of the present paper is on the
construction of unconditionally stable algorithms we relegate
a presentation of these technical but for applications important extensions
to future publications.

%
%-------------------------------------------------------------
%
\section{Data analysis}\label{sec5}
%
%-------------------------------------------------------------
%

Time-domain algorithms obviously yield the time development of the EM fields.
The scattering (transmission) of the EM fields from (through) objects is one
of the main applications of this technique~\cite{Taflove}.
One approach is to prepare an initial state $\Psi(0)$ of the EM fields,
propagate the fields in time for a number of time-steps and analyse the
scattered and/or transmitted fields.
Another, more realistic, approach is to use a current source
${\mathbf{J}}(t)={\mathbf{J}}(\br,t)$.
Instead of Eq.~(\ref{TDME}) we have
\begin{equation}
\dd{t}\Psi(t)={\mathcal H}\Psi(t) - {\mathcal J}(t),
\label{TDMEJ}
\end{equation}
where $\Psi(t)=(\bX(t),\bY(t))^T$ and
${\mathcal J}(t)=(0,{\mathbf{J}}(t))^T$ represents the source term.
The formal solution of Eq.~(\ref{TDMEJ}) is given by
\begin{equation}
\Psi(t) = e^{t{\mathcal H}}\Psi(0)-\int_0^t
e^{(t-u){\mathcal H}}{\mathcal J}(u)\,du,
\label{formalJ}
\end{equation}
showing that we can simply re-use one of the unconditionally stable
algorithms to compute the second term in Eq.~(\ref{formalJ}).
In practice, for a time-step $\tau$, we update $\Psi(t)$ according to
\begin{equation}
\Psi(t+\tau) = e^{\tau{\mathcal H}}\Psi(t)-\int_t^{t+\tau}
e^{(t+\tau-u){\mathcal H}}{\mathcal J}(u)\,du.
\label{Psitau}
\end{equation}
A standard quadrature formula can be used to compute the integral over
$u$~\cite{Michielsen98}. When a current source is present
we take as the initial condition $\Psi(0)=0$.

Time-domain algorithms can also be used to compute the eigenvalues of $H$, the discretized
form of ${\mathcal H}$.
In general $H$ is a (very) large matrix, usually too large to be stored in
memory.
If only a few, well-separated eigenvalues of $H$ are required sparse-matrix
techniques can be used to compute these eigenvalues~\cite{GOLUB,WILKINSON}.
However, if one is interested in global features of the distribution of
eigenvalues, i.e. if we want to determine {\sl all} eigenvalues,
time-domain algorithms offer several advantages.
In fact they are at the heart of so-called ``fast'' algorithms to compute
the density of states (DOS) and other related
quantities~\cite{ALBEN,FEIT82,HANS89,KAWA96,OHT97,IITAKA99}.
The basic idea of this approach was laid out by Alben et al.~\cite{ALBEN}
who used it to compute the DOS of models for one electron moving in a disordered
alloy.

Denoting the (unknown) eigenvalues and (unknown) eigenvectors of $H$ by
$iE_j$ and $\phi_j$ respectively, we have
\begin{eqnarray}
f(t)&\equiv&
\frac{\langle\Psi(0)|\Psi(t)\rangle}{\langle\Psi(0)|\Psi(0)\rangle}
=\frac{\langle\Psi(0)|e^{tH}\Psi(0)\rangle}{\langle\Psi(0)|\Psi(0)\rangle}
=\sum_{i=1}^K
e^{itE_i}
\frac{|\langle\Psi(0)|\phi_i\rangle|^2}{\langle\Psi(0)|\Psi(0)\rangle}\,,
\label{ft}
\end{eqnarray}
where $K$ ($K=n$ for 1D, $K=3n_x n_y/4$ for 2D, $K=3n_xn_yn_z/4$ for 3D)
is the dimension of the vector space on which $H$ acts.
From Eq.~(\ref{ft}) it follows immediately that the Fourier transform of
$f(t)$ contains the information on all eigenvalues for which
$|\langle\Psi(0)|\phi_i\rangle|>0$.
Using independent random numbers to initialize the elements of $\Psi(0)$,
it can be shown that the density of states ${{\mathcal D}}(\omega)$ is
given by \cite{Hams00}
\begin{eqnarray}
{{\mathcal D}}(\omega)=a \int_{-\infty}^{+\infty} e^{-i\omega t}
\overline{f(t)}\, dt,
\label{DOS}
\end{eqnarray}
where $a$ is an irrelevant constant factor and $\overline{f(t)}$ is the
average of $f(t)$ over different realizations of the random initial state.
It is often expedient to consider, in addition to ${{\mathcal D}}(\omega)$,
the integrated density of states
\begin{equation}
N(\omega)=\int_{-\infty}^{\omega}{\mathcal D}(u)\,du\,.
\end{equation}
The statistical error on $f(t)$ vanishes as $1/\sqrt{SK}$ where $S$ is the
number of statistically independent samples of $\Psi(0)$ \cite{Hams00}.
The fact that the statistical error decreases with the number of lattice
points $K/2$ gives a tremendous boost to the efficiency of the method.

The information on the eigenvalues of $H$, obtained through the use of a
time-domain method is intimately related to the unconditional stability of
the latter.
As $f(t)$ is band-limited, with frequencies $E_j$ in the interval
$[-\Vert H\Vert,\Vert H\Vert]$ (where $\Vert H\Vert$ denotes the largest
eigenvalue of H in absolute value), it follows from Nyquist's sampling
theorem that it is sufficient to sample $f(t)$ at regular intervals
$\Delta t = \pi/\Vert H\Vert$.
If $N$ denotes the number of data points used to sample $f(t)$, frequencies
$E_j$ that differ less than $\Delta E=\pi/N\Delta t$ are
indistinguishable (although they will all contribute to the DOS).
Extending the length of the time integration by a factor of two increases
the resolution in frequency by a factor of two.
This is a rather efficient and flexible procedure to trade accuracy for
computational resources.
One may object that by integrating the TDME over longer and longer times
(larger and larger $N$) the error on $\Psi(t)$ will increase, possibly
leading to no gain in accuracy at all.
However, it has been shown rigorously~\cite{Hams00} that the error on the
eigenvalues of $H$ vanishes as $\tau^2/ N$ if one uses unconditionally
stable algorithms based on the second-order Suzuki product-formula.
The proof given in Ref.~\cite{Hams00} applies to the fourth-order algorithm
$U_4(\tau)$ as well, the exponent of $\tau$ being four instead of two.

In some cases the underlying differential equations and boundary conditions
only specify the solution up to a non-zero constant.
%This is the case for the Maxwell equations in 2D and 3D.
In the calculation of the DOS, the presence of such a constant contribution
shows up as a peak at zero frequency.
In principle this peak can be removed by modifying the random initial state
but as the origin of this irrelevant artifact is understood, there is little
reason of doing this.

Summarizing: Solving the TDME by the $k$'th-order Suzuki product-formula
algorithm $U_k(\tau)$ guarantees that the accuracy with which the
eigenvalues of $H$ can be determined vanishes as $\tau^k/N$, where $N$ is the
number of points used to sample $f(t)$ (see Eq.~(\ref{ft})).

%
%-------------------------------------------------------------
%
\section{Simulation results}\label{sec6}
%
%-------------------------------------------------------------
%

In this section we present simulation results for several physical systems
which we selected as examples to test our algorithms.
For numerical purposes it is expedient to use dimensionless quantities.
We will denote the unit of length by $\lambda$ and take the velocity of light
in vacuum, $c$, as the unit of velocity.
Then time and frequency are measured in units of $\lambda/c$ and $c/\lambda$,
respectively.
The permittivity $\varepsilon$ and permeability $\mu$ are measured in units of their corresponding
values in vacuum.

%%%%%%%%%%%%%%
%
\subsection{One dimension}\label{subsec61}
%
%%%%%%%%%%%%%%

\begin{figure}[t]
\setlength{\unitlength}{1cm}
\begin{picture}(9.2,6.6)
\put(0.5,0){\epsfig{file=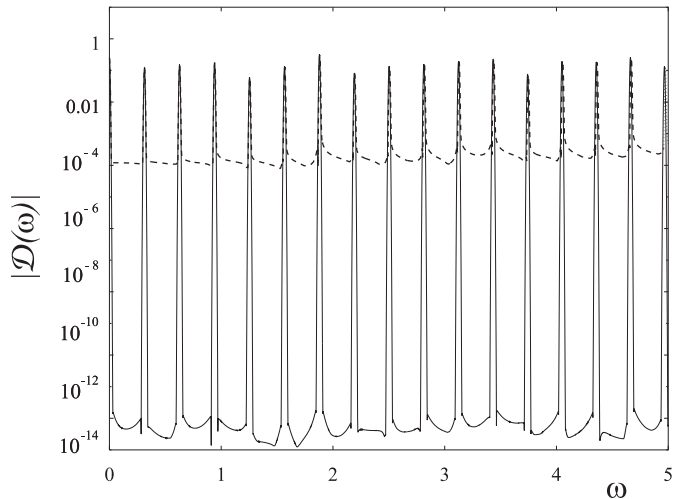}}
\end{picture}
\caption{Density of states of a one-dimensional cavity of length $L=10$.
Solid line: $U_2(\tau)$ algorithm; dashed line: standard Yee algorithm.}
\label{fig:spec1D}
\end{figure}

\begin{figure}[t]
\setlength{\unitlength}{1cm}
\begin{picture}(9.2,6.7)
\put(0.5,0){\epsfig{file=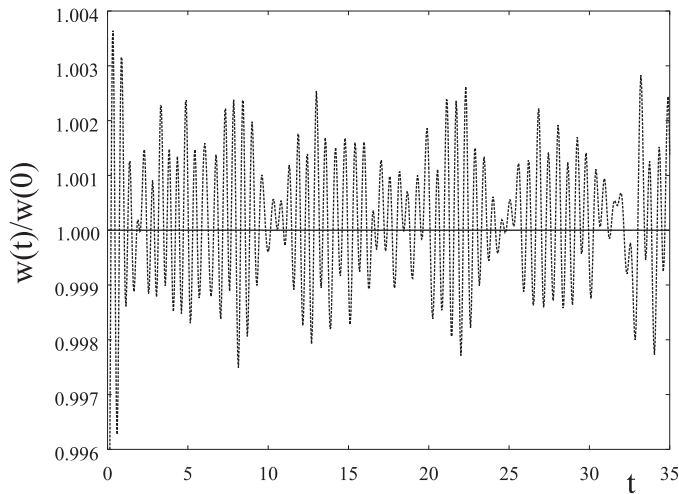}}
\end{picture}
\caption{Normalized total energy density as a function of time for the same
physical system as in Fig.~\ref{fig:spec1D}.
Solid line: $U_2(\tau)$; dashed line: standard Yee algorithm.}
\label{fig:normYee}
\end{figure}

Let us first consider an empty, one-dimensional cavity with constant
permeability $\mu=1$ and constant permittivity
$\varepsilon=1$.
The eigenfrequencies for a system of length $L$ are given by~\cite{BornWolf}
\begin{equation}
\omega_k=\frac{\pi k}{L}\,,
\label{omegak}
\end{equation}
where $k=0,1,2,\ldots$ labels the different EM modes.
In Fig.~\ref{fig:spec1D} we show the density of states $|{\cal D}(\omega)|$
obtained according to the procedure described in Sec.~\ref{sec5}, using the
second-order algorithm $U_2(\tau)$ (solid line) and the standard Yee
algorithm~\cite{Yee66,Taflove} (dashed line).
\noindent
In both calculations the lattice spacing $\delta=0.1$ and the time step
$\tau=0.01$.
Each curve shown in Fig.~\ref{fig:spec1D} is the average of $S=10$
statistically independent runs, taking $N=16384$ samples of $f(t)$
(see Eq.~(\ref{ft})) at time intervals $\Delta t=0.1$.
The peaks in Fig.~\ref{fig:spec1D} correspond to the exact frequencies
(see Eq.~(\ref{omegak})) of the 1D cavity.
The background signal produced by the Yee algorithm~\cite{Yee66,Taflove}
is at least eight orders of magnitude larger than that generated by
$U_2(\tau)$.
The time-step operator of the Yee algorithm is not an orthogonal matrix
and hence its eigenvalues do not necessarily lie on the unit circle.
This is related to the fact that the Yee algorithm is conditionally
stable~\cite{Taflove} and leads to fluctuations in the energy density
$w(t)$, as illustrated by the dashed line in Fig.~\ref{fig:normYee},
and to negative values of the Fourier transform of $f(t)$
(which is the reason why Fig.~\ref{fig:spec1D}
shows $|{\cal D}(\omega)|$ instead of ${\cal D}(\omega)$).

In contrast, the unconditional stability of algorithms based on the Suzuki
product-formula implies that $w(t)$ is constant in time.
The solid line in Fig.~\ref{fig:normYee} shows that this is indeed the case.

\begin{figure}[t]
\setlength{\unitlength}{1cm}
\begin{picture}(6.8,2.0)
\put(0.5,0){\epsfig{file=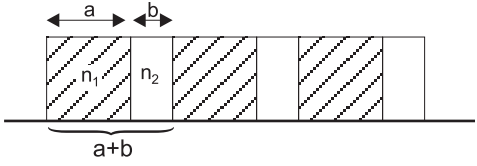}}
\end{picture}
\caption{Structure of a one-dimensional stack of dielectric material.}
\label{fig:fig4}
\end{figure}

\begin{figure}[t]
\setlength{\unitlength}{1cm}
\begin{picture}(8.9,6.4)
\put(0.5,0){\epsfig{file=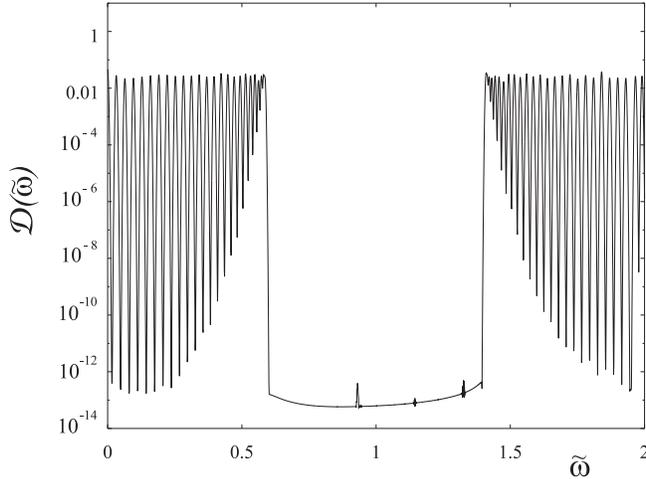}}
\end{picture}
\caption{Density of states, as obtained by the $U_2(\tau)$ algorithm,
of a quarter-wave stack of length $L=24.9$
and with parameters $n_1=1$, $n_2=4$, $a=0.8$, $b=0.2$
(see Fig.\ref{fig:fig4}) as a function of the rescaled frequency
$\tilde{\omega}=\omega/\omega_0$.}
\label{fig:QWS1_spec}
\end{figure}

As a second example we consider a one-dimensional stack of dielectric
material, schematically shown in Fig.~\ref{fig:fig4}.
The material indices of refraction, denoted by $n_1$ and $n_2$, give rise to
a spatially varying permittivity
\begin{equation}
\ve(x)=\left\{ \begin{array}{l} n_1^2, {\rm\ if\ } x\,{\rm mod }
(a+b) \leq a \\ n_2^2, {\rm\ if\ } x\,{\rm mod }(a+b) > a
\end{array}\right..
\end{equation}
\noindent
In particular, we consider a structure that is known as the quarter-wave
stack and is characterized by the relation
\begin{equation}
n_1 a = n_2 b,
\end{equation}
such that the length of the optical path in the two layers is the same.
The density of states exhibits a gap centered
around the midgap frequency~\cite{BornWolf,Bendickson96}
\begin{equation}
\omega_0=\frac{\pi}{2n_1 a}.
\end{equation}
In Fig.~\ref{fig:QWS1_spec} we show the density of states
${\cal D}(\omega)$ as obtained by $U_2(\tau)$.
Also for these calculations the lattice spacing $\delta=0.1$ and the time
step $\tau=0.01$.
Each curve in Fig.~\ref{fig:QWS1_spec} is the average of $S=100$
statistically independent runs, taking $N=16384$ samples of $f(t)$ (see
Eq.~(\ref{ft})) at time intervals $\Delta t=0.1$.
Note that the system length $L$ and the (odd) number of lattice points have
to be chosen judisciously, otherwise the spectrum will exhibit artifacts
(impurity states due to one extra grid point).

\begin{figure}[t]
\setlength{\unitlength}{1cm}
\begin{picture}(8.9,6.4)
\put(0.5,0){\epsfig{file=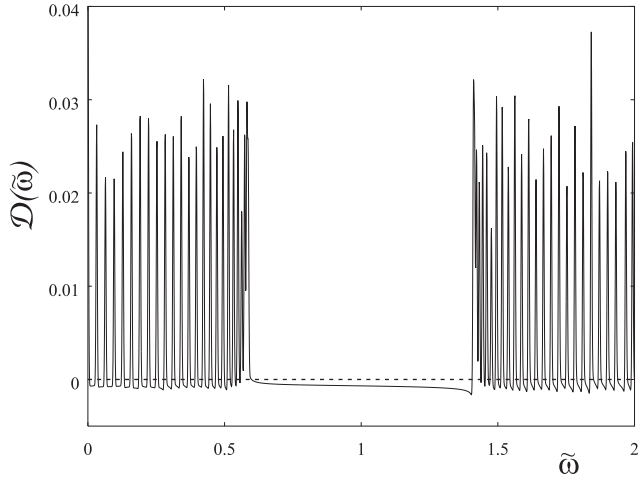}}
\end{picture}
\caption{Density of states, as obtained by
the standard Yee algorithm, of a quarter-wave stack of length $L=24.9$
and with parameters $n_1=1$, $n_2=4$, $a=0.8$, $b=0.2$
(see Fig.\ref{fig:fig4}) as a function of the rescaled frequency
$\tilde{\omega}=\omega/\omega_0$.
Note the difference in the vertical scale between Figs.\ref{fig:QWS1_spec} and \ref{fig:QWS2_spec}.}
\label{fig:QWS2_spec}
\end{figure}

\begin{figure}[t]
\setlength{\unitlength}{1cm}
\begin{picture}(9,6.8)
\put(0.5,0){\epsfig{file=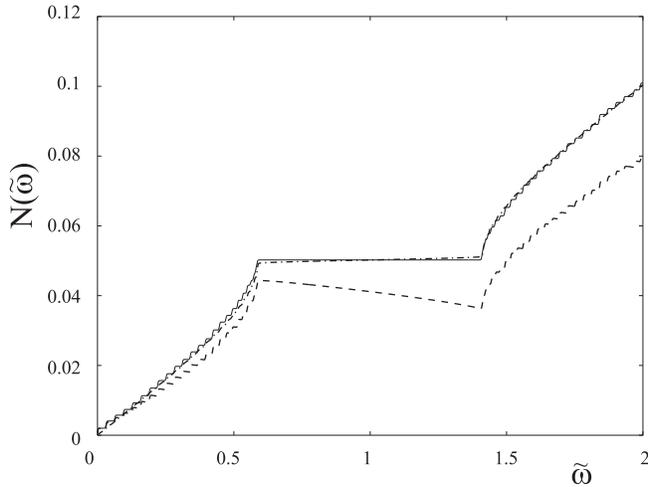}}
\end{picture}
\caption{Integrated density of states as function of the rescaled frequency
$\tilde\omega=\omega/\omega_0$,
for the same system as in Figs.\ref{fig:QWS1_spec} and \ref{fig:QWS2_spec}.
Solid line: $U_2(\tau)$ algorithm;
dashed line: standard Yee algorithm;
dashed-dotted line: analytical result. %Ref.{\cite{Bendickson96}}.
}
\label{fig:QWS1_idos}
\end{figure}

In Fig.~\ref{fig:QWS2_spec} we show the density of states
${\cal D}(\omega)$ as obtained by the Yee algorithm.
Note that the spectral weight can take negative values,
an unphysical feature that is a manifestation of the
fact that the energy of the EM field is not conserved.
In Fig.~\ref{fig:QWS1_idos} we present the integrated density of states
$N(\omega)$ for both the $U_2(\tau)$ and Yee algorithm.
The result of the $U_2(\tau)$ algorithm is in excellent agreement
with the analytical calculation ~\cite{Bendickson96}.

%%%%%%%%%%%%%%%%
%
\subsection{Two dimensions}\label{subsec62}
%
%%%%%%%%%%%%%%%

A photonic bandgap (PBG) material prohibits the propagation of EM fields
in a range of frequencies that is characteristic for its
structure~\cite{Yabl87}.
A PBG is called absolute if it exists for any wave vector of the EM fields.
The most common method used to compute a PBG employs a plane-wave expansion
to solve the time independent Maxwell equations (see e.g. \cite{Ho90}).
This kind of PBG calculation requires a Fourier transform of the unit cell of
the dielectric structure which is for simplicity considered to be periodic.

\begin{figure}[t]
\setlength{\unitlength}{1cm}
\begin{picture}(9.1,6.6)
\put(0.5,0){\epsfig{file=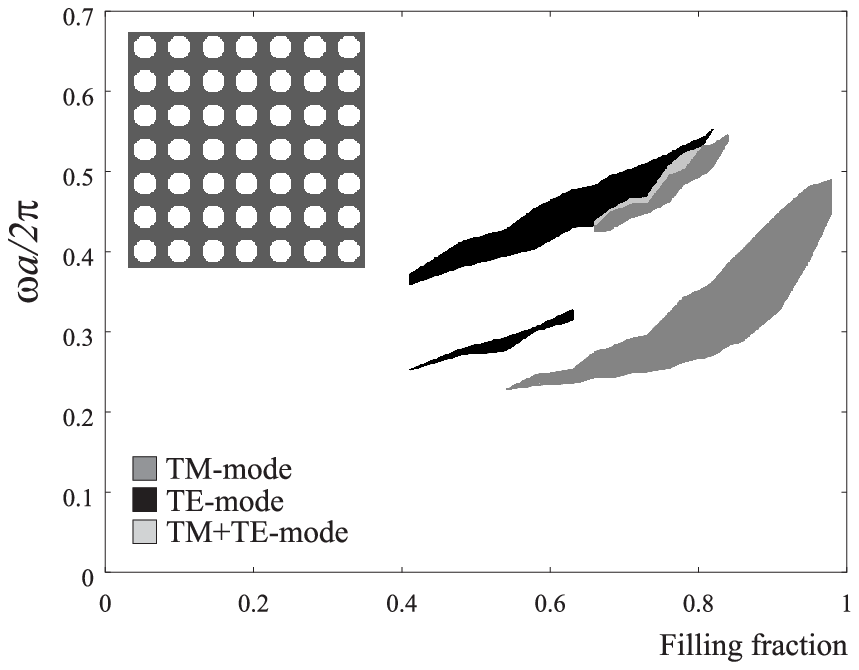}}
\end{picture}
\caption{Absolute photonic bandgaps of a
dielectric material ($\varepsilon=11.4$) pierced by air-filled cylinders.
The largest overlap of the TM- and TE-mode gaps occurs at a filling fraction
of approximately 0.77.}
\label{fig:bandgaps}
\end{figure}

With our time-domain algorithm the existence of a PBG can be demonstrated
with relative ease.
It suffices to compute the spectrum of such a dielectric structure with a
random initial state.
If the spectrum is gapless there is no need to make additional runs.
If there is a signature of a gap, it can be confirmed and refined by making
more runs.
As an example we consider a system consisting of a dielectric material pierced
by air-filled cylinders~\cite{Anderson96}.
The geometry is taken to be a square parallelepiped of size $L=45.1$ that is
infinitely extended in the $z$-direction
and hence is effectively two-dimensional.
In Fig.~\ref{fig:bandgaps} we present the results for the PBGs which we
obtained for both the transverse magnetic (TM) and transverse electric (TE)
modes as a function of the filling fraction.
The data have been generated by means of the algorithm $U_4(\tau)$ with a mesh
size $\delta=0.1$ and a time step $\tau=0.1$.
To compute the DOS we used $N=32768$ samples of $f(t)$ at time intervals
$\Delta t=0.1$.
Only a single random initial state for the EM fields has been used.
Replacing the free-end boundary conditions Eq.~(\ref{boundcond}) by periodic
boundary conditions (results not shown) only leads to minor changes in the
locations of PBGs.
The results shown in Fig.~\ref{fig:bandgaps} are in good agreement with
those presented in Ref.~\cite{Anderson96}.

\begin{figure}[t]
\setlength{\unitlength}{1cm}
\begin{picture}(8.3,7.9)
\put(0.5,0.2){\epsfig{file=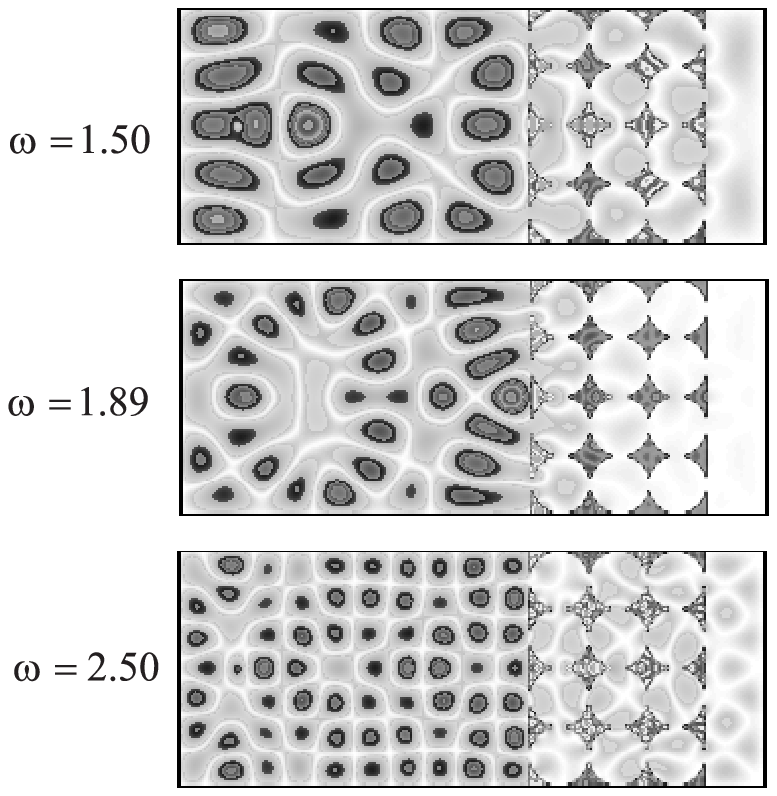}}
\end{picture}
\caption{Snapshot of the intensity $E_z^2$ at $t=102.4$.
Dimensions of the system: $30\times12.1$;
point source located at (6,6) (see Fig.\ref{fig:2DTM}), emitting radiation with frequency $\omega$.
} \label{fig:2Dtimedep}
\end{figure}

\begin{figure}[t]
\setlength{\unitlength}{1cm}
\begin{picture}(9.1,9.2)
\put(0.5,0){\epsfig{file=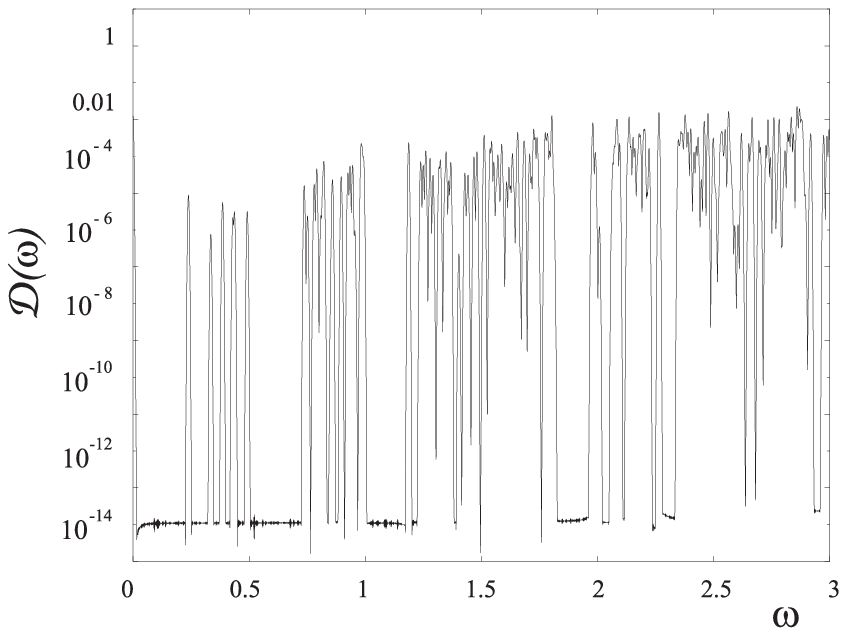}}
\end{picture}
\caption{Density of states of a sample of photoning band-gap material used in
Fig.\ref{fig:2Dtimedep}.
Size of the sample: $9.1\times12.1$; filling factor: 0.77.}
\label{fig:2Dspec}
\end{figure}

In Fig.~\ref{fig:2Dtimedep} we study the propagation of time-dependent EM
fields through the above described PBG material consisting of twelve unit
cells.
\noindent
The PBG material is placed in a cavity which contains a point source (located
to the left of the PBG material) that emits radiation with frequency $\omega$.
The TDME were solved by the $U_2(\tau)$ algorithm with $\delta=0.1$ and
$\tau=0.01$ in the presence of a current source according to
Eq.~(\ref{Psitau}).
The snapshots show the absolute intensity $E_z^2$ of the TM-mode at
$t=102.4$.
The computed DOS of the PBG material is given in Fig.~\ref{fig:2Dspec}.
We used the $U_4(\tau)$ algorithm with $\delta=0.1$, $\tau=0.1$, and took
$N=32768$ samples of $f(t)$ at time intervals $\Delta t=0.1$ in this
computation.
\noindent
The presence or absence of gaps in the DOS leads to qualitative changes in
the transmitted (and reflected) intensities.
Since a gap is present in the DOS at $\omega=1.89$, radiation with this
frequency does not easily propagate through the (thin slice of) PBG material.
On the other hand, the DOS has no gaps at $\omega=1.50$ and $\omega=2.50$,
so that propagation of EM fields through the PBG material should be possible,
as is indeed confirmed by Fig.~\ref{fig:2Dtimedep}.

%%%%%%%%%%%%%%%
%
\subsection{Three dimensions}\label{subsec63}
%
%%%%%%%%%%%%%%%

We first compute the modes of a simple cubic cavity of size $L$.
The eigenfrequencies are given by~\cite{BornWolf}
\begin{equation}
\omega_{klm}=\pi L^{-1}\sqrt{k^2+l^2+m^2}\,,
\end{equation}
where $k$, $l$, and $m$ are non-negative integers.
Eigenmodes corresponding to $(k,0,0)$, $(0,l,0)$ and $(0,0,m)$ are
incompatible with the boundary conditions Eq.~(\ref{boundcond}).
In Tab.~\ref{tbl:table3D} we present the frequencies of the five lowest
eigenmodes of a cubic cavity for $L=5$.
\begin{table}[ht]
  \begin{center}
    \begin{tabular}{ccc}
$(k,l,m)$&$\omega_{klm}$&Simulation\\ \hline 1,1,0 & 0.889 & 0.889
\\ 1,1,1 & 1.088 & 1.089 \\ 2,1,0 & 1.405 & 1.404 \\ 2,1,1 &
1.539 & 1.534 \\ 2,2,0 & 1.777 & 1.771 \\
    \end{tabular}
    \end{center}
  \caption{Frequencies of the eigenmodes of a cubic cavity of size
$L=5$. Simulation: values determined from DOS data generated using
$U_2(\tau)$ with parameter values $\delta=0.2$, $\tau=0.01$, $N=4096$
and $\Delta t=0.1$.}
\label{tbl:table3D}
\end{table}
The agreement with the theoretical values is satisfactory.
Note that as the frequency increases, the deviation from the
exact result increases. This is due to the (second-order) finite-difference
representation of the spatial derivatives on the grid and is a
manifestation of the numerical dispersion mentioned earlier.

\begin{figure}[t]
\setlength{\unitlength}{1cm}
\begin{picture}(12.,12.)
\put(0.5,.5){\epsfig{file=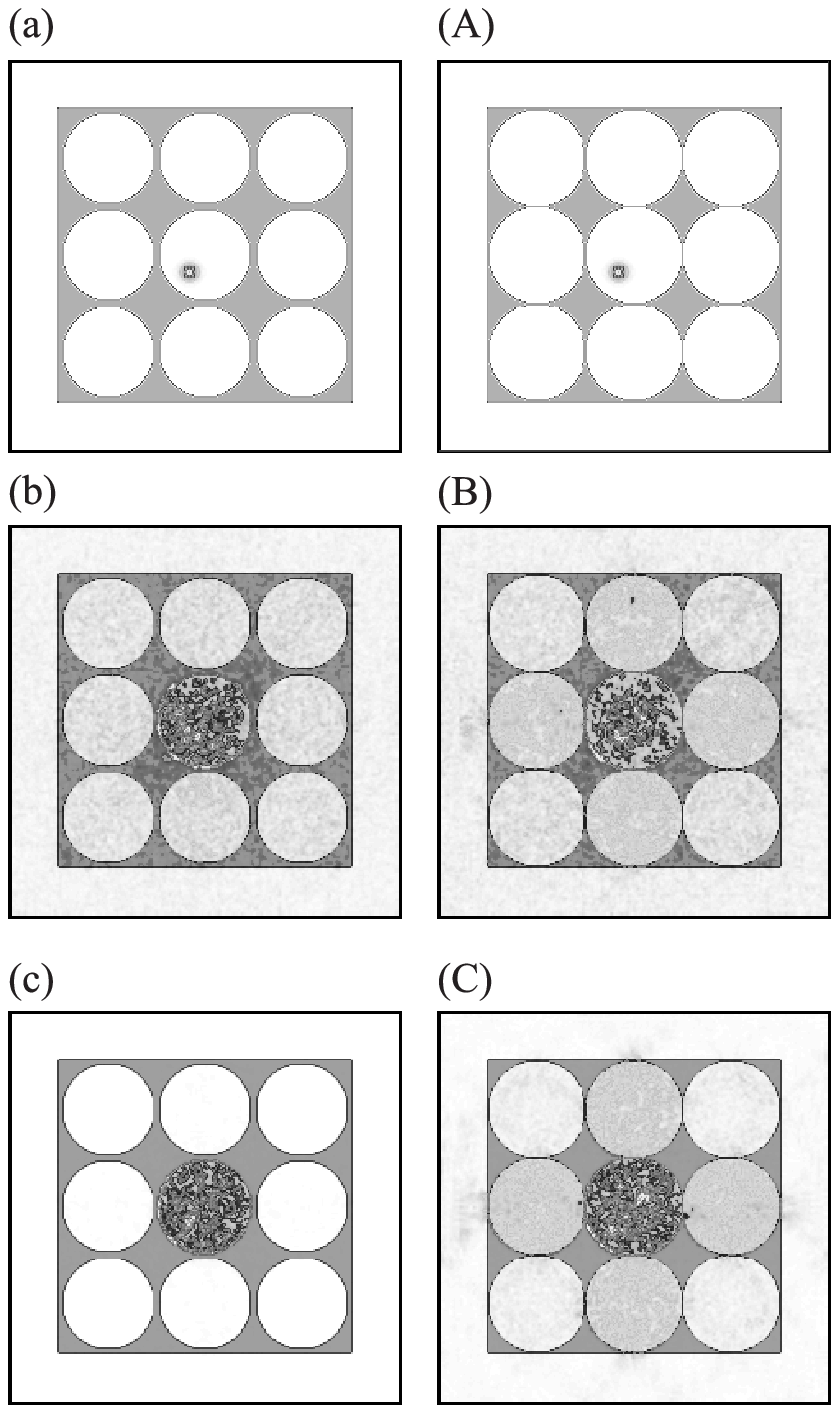}}
\end{picture}
\caption{Intensity of the EM-fields in and outside a cubic sample of dielectric material
containing $3\times3\times3$ spherical voids, placed in an empty cavity.
The size of the cavity is $12.1\times12.1\times12.1$,
the size of the sample is $9.1\times9.1\times9.1$,
the source is located at $(5.6,5.5,6)$,
the slices shown are at $z=6$, the fourth-order algorithm $U_4(\tau)$
was used with a time step $\tau=0.075$ and mesh size $\delta=0.1$.
(a): Intensity at $t=0.3$. The radius of the empty spheres is $1.4$.
(b): Intensity at $t=384$. Same system as in (a).
The permittivity of the dielectric material $\varepsilon=1.5$.
(c): Intensity at $t=384$. Same system as in (b).
The permittivity of the dielectric material $\varepsilon=5$.
(A): Intensity at $t=0.3$. The radius of the empty spheres is $1.5$.
(B): Intensity at $t=384$. Same system as in (A).
The permittivity of the dielectric material $\varepsilon=1.5$.
(C): Intensity at $t=384$. Same system as in (B).
The permittivity of the dielectric material $\varepsilon=5$.
}
\label{fig:PGB3D}
\end{figure}

As a second 3D example we consider the emission of the EM radiation
from a point source located inside dielectric material containing spherical voids.
A projection of the material onto the $x-y$ (or $y-z$ or $x-z$) plane is
shown in the top panels of Fig.\ref{fig:PGB3D}.
A point source is placed inside the center void to mimic an atom or molecule that emits a photon.
The remaining panels in Fig.\ref{fig:PGB3D} show snapshots of the light intensity after
an elapsed time $t=384$, for different sizes of the voids and different values
of the permittivity. If the latter is larger than 5, the images no longer
depend on the value of the permittivity (results not shown).
For $\varepsilon=1.5$ (panels (b) and (B)) the EM field easily propagates through
the sample and leaves the sample from all sides. This is not the case
for $\varepsilon=5$: Radiation can leave the sample only at those locations where
there is very little or no dielectric material left. In other words, light emerges
from the sample in well-defined directions only.
Clearly, much more work is necessary to study the propagation of EM radiation in this system
as a function of the material parameters, the system size and the frequency of the
emitted photons.

%

%
%-------------------------------------------------------------
%
\section{Conclusion}\label{sec7}
%
%-------------------------------------------------------------
%

We have introduced a new family of algorithms to solve the time-dependent Maxwell
equations. Salient features of these algorithms are:

\begin{itemize}
\item rigorously provable unconditional stability for one-, two- and three-dimensional
systems with spatially varying permittivity and permeability,
\item the use of a real-space (Yee-like) grid,
\item the order of accuracy in the time step can be systematically increased without affecting the
unconditional stability (in this paper we limited ourselves to second and fourth-order schemes)
\item the exact conservation of the energy density of the electromagnetic field
\item easy to implement in practice
\end{itemize}

We have presented results for the density of states of simple cavities
and photonic bandgap materials. We demonstrated that mathematical properties
of the algorithms are such that they can be used to compute the density of states
with very good accuracy (limited by the accuracy of the spatial discretization used).
We gave some illustrative examples of scattering of waves by photonic bandgap
systems. These examples also served to show that our algorithms reproduce known results.
The first feature opens up possibilities for applications to left-handed materials~\cite{Veselago68,Shelby01}.
We intend to report on this subject in the near future.

Although we believe there is little room to improve upon
the time-integration scheme itself (except for using higher-order product-formulae),
for some applications it will be necessary to use a better spatial discretization than
the most simple one employed in this paper. There is no fundamental problem to
extend our approach in this direction and we will report on this issue
and its impact on the numerical dispersion in a future publication.

The rigorous unconditional stability of the algorithms that we proposed
in this paper is a direct consequence of adopting a Suzuki-product-formula
approach that preserves the fundamental symmetries of the physical system.
In view of the generic character of this methodology, the approach
pursued in the present paper should be useful for constructing
unconditionally stable algorithms that solve the equations for e.g.
sound, seismic and elastic waves as well.

%
%-------------------------------------------------------------
%
\section*{Acknowledgements}
%
%-------------------------------------------------------------
%
We thank K. Michielsen for a critical reading of the manuscript and W. Bruns for helpful discussions.
This work is partially supported by the Dutch `Stichting Nationale Computer Faciliteiten'
(NCF).

\begin{thebibliography}{99}
\bibitem{COMMENT1}
This is due to the specific split-up adopted in \cite{Harsh00}
and not an intrinsic property of the spectral-domain approach.
\end{thebibliography}
\end{document}